\documentclass[aps,prc,twocolumn,superscriptaddress,altaffilletter]{revtex4-1}
\usepackage{pifont}
\usepackage{color}
\usepackage{amsmath}
\usepackage[makeroom]{cancel}
\usepackage{amssymb}
\usepackage{natbib}
\usepackage{graphicx}
\usepackage{lineno}
\usepackage[T1]{fontenc}
\usepackage[utf8]{inputenc}
\usepackage[margin=1in]{geometry}
\usepackage{hyperref}

\usepackage{type1cm}
\usepackage{eso-pic}
\usepackage{threeparttable}
\usepackage{dcolumn}
\usepackage{xhfill}

\newcommand{\krm}{$^{83}$Kr$^m$}

\begin{document}
\title{Shakeup and shakeoff satellite structure in the electron spectrum of $^{83}$Kr$^m$ }

\author{R.G.~Hamish Robertson}\email[]{rghr@uw.edu}
\affiliation{Center for Experimental Nuclear Physics and Astrophysics and Department of Physics, University of Washington, Seattle, WA 98195} 

\author{Vedantha Venkatapathy}
\altaffiliation{Present address: Dept. of Physics, Massachusetts Institute of Technology, Cambridge MA 02139.}\affiliation{Lake Washington School District, Redmond, WA 98073}


\begin{abstract}
The isotope $^{83}$Kr$^m$, a 1.8-hr isomer of stable $^{83}$Kr, has become a standard for the calibration of tritium beta decay experiments to determine neutrino mass. It is also widely used as a low-energy electron source for the calibration of dark-matter experiments.  The nominally monoenergetic internal conversion lines are accompanied by shakeup and shakeoff satellites that modify the line shape.  We draw on theoretical and experimental information to derive a quantitative description of the satellite spectrum of the K-conversion line.  

\end{abstract}
\keywords{neutrino mass, internal conversion, shakeup, shakeoff, electron spectroscopy}
\maketitle

\section{Introduction}

Sensitive experiments to measure the mass of the neutrino are based on the beta decay of molecular tritium \cite{robertson:1991aa,Stoeffl:1995wm,Kraus:2004zw,Belesev:2008zz,Aseev:2012zz,PhysRevLett.123.221802,Asner:2014cwa}.    In those experiments, without exception, the isotope \krm\  is used to measure and verify the instrumental response.  The isotope is also used in the calibration of dark matter experiments with xenon targets \cite{Akerib:2017eql,Aprile:2018dbl,PhysRevC.80.045809}.  The 1.8-hr isomer is produced conveniently via the beta decay of the longer-lived 86-d $^{83}$Rb, and it  decays to the stable ground state via two sequential  transitions,  32 keV and 9.4 keV.  The transitions are internally converted to a large degree and produce a complex spectrum of conversion and Auger electrons. The widths of the conversion lines are determined by the lifetimes of the vacancies created by the conversion, and are typically a few eV.  The K-conversion line at 17.8 keV (the `K-32' line) is particularly useful as it has a narrow natural width of 2.7 eV and is close to the endpoint of the tritium spectrum at 18.6 keV.   

The conversion lines are modified by shakeup and shakeoff processes, which produce satellite structures on the low-energy wings of the lines.  The importance of a quantitative description of the satellite spectrum was recognized even with the first use of  \krm \cite{robertson:1991aa} in a tritium experiment. A weak, broad structure about 100 eV below the 17.8-keV conversion line was noted in the data, which was not consistent with the expected instrumental response.  Any unaccounted contribution $\Delta \sigma^2$ to the variance of the instrumental response contributes directly to the neutrino mass $m_\nu^2$ as expressed by the approximate relationship \cite{Robertson:1989qj}, 
\begin{eqnarray}
\Delta m_\nu^2&\simeq& -2 \Delta\sigma^2. \label{eq:variance}
\end{eqnarray}
 It was therefore  important to identify the observed satellite structure as either an instrumental effect or a property of the calibration line.    
The theoretical work of Carlson and Nestor \cite{PhysRevA.8.2887} predicted the presence of shakeup structure there, and an experiment to measure it  by photoelectron spectroscopy was carried out at the Stanford  Synchrotron Radiation Laboratory's PEP X-ray source \cite{PhysRevLett.67.2291}.  That experiment not only confirmed the presence of the shakeup structure, it gave quantitative validation for the assumed equivalence of internal conversion and photoelectron ejection in the electron spectra generated.  To help interpret the experimental data, a Relativistic Dirac-Fock (RDF) theoretical calculation of the two-vacancy process was carried out, giving the energies and intensities of the satellite shake structure out to 300 eV from the core K electron line.  

The problem addressed in the present work is that the theoretical results in \cite{PhysRevLett.67.2291} are in only qualitative agreement with experiment.  The experimental results themselves are reliable as evidenced by the close agreement between internal conversion and photoionization, but they were taken with modest resolution.  Consequently, there is a lack of quantitative spectral information that can be applied both to the calibration of modern high-resolution instruments, and to advancing the theory of shakeup and shakeoff structure in Kr and other complex atoms.

\section{Conversion-Line Spectrum}

     A major experimental study of the conversion line spectrum of \krm\  was made by Picard {\em et al.} using a frozen source and the Mainz spectrometer \cite{picard:1992aa}.  They measured the energies and widths of the conversion lines and used tabulated electron binding energies to deduce the transition energies.  Since then, more precise measurements of the calibration standards have improved the accuracy of the energies. 
     
     A comprehensive summary of the energies of the conversion lines is given by V\'{e}nos {\it et al.} \cite{Venos:2018tnw}. Table \ref{tab:lines} displays the electron energies, intensities, and line widths for internal conversion of the 32-keV transition. 
\begin{table}[htb]
\caption{Conversion electron lines from V\'{e}nos {\em et al.} \cite{Venos:2018tnw}.}
\medskip
\begin{center}
\begin{tabular}{lrrrr} 
\hline
\hline
Line&Energy (eV)&Intensity (\%)& Width (eV)&\\
\hline&&&&\\
$1s_{1/2}K $&17824.2(5)&24.8(5)&2.70(6)&\\
$2s_{1/2}L_1$&30226.8(9)&1.56(2)&3.75(93)&\\
$2p_{1/2}L_{2}$&30419.5(5)&24.3(3)&1.165(69)&\\
$2p_{3/2}L_{3}$&30472.2(5)&37.8(5)&1.108(13)&\\
$3s_{1/2}M_1$&31858.7(6)&0.249(4)&3.5(4)&\\
$3p_{1/2}M_{2}$&31929.3(5)&4.02(6)&1.230(61)&\\
$3p_{3/2}M_{3}$&31936.9(5)&6.24(9)&1.322(18)&\\
$3d_{3/2}M_{4}$&32056.4(5)&0.0628(9)&0.07(2)&\\
$3d_{5/2}M_{5}$&32057.6(5)&0.0884(12)&0.07(2)&\\
$4s_{1/2}N_1$&32123.9(5)&0.0255(4)&0.40(4)&\\
$4p_{1/2}N_2$&32136.7(5)&0.300(4)&0&\\
$4p_{3/2}N_3$&32137.4(5)&0.457(6)&0&\\
\hline
\end{tabular}
\end{center}
\label{tab:lines}
\end{table}

In recent unpublished results \cite{Altenmuller:2019ddl} on the Kr spectrum from KATRIN,  the width of the K-32 line is given as $2.774\pm0.011_{\rm stat}\pm0.005_{\rm syst}$ eV.  From earlier KATRIN measurements with a solid source \cite{Arenz2018} a slightly smaller value, 2.70 eV, may be derived.  Our work, which is not particularly sensitive to this quantity, adopts the widths given by V\'{e}nos {\em et al.} \cite{Venos:2018tnw}.

\section{Shakeup and shakeoff}

Ejection of a conversion electron or a photoelectron sometimes creates more than a single vacancy in the daughter atom, because the atomic wave functions of the electron states in parent and daughter do not overlap perfectly.  These additional vacancies tend to occur in the outer shells when a core-shell electron is ejected, and they lead to lower-energy satellite structures adjacent to the core conversion line.  Those satellites should be taken into account when deriving the instrumental resolution from the profile of a conversion line.  

As has been mentioned, the first case where this arose was the Los Alamos (LANL) tritium beta decay experiment \cite{robertson:1991aa}.  The resolution function was determined from the Kr K-32 line shape, which in turn was measured by photoelectron spectroscopy at the SSRL-PEP synchrotron radiation source \cite{PhysRevLett.67.2291}.  The satellite spectrum was measured to an energy 300 eV below the core line.  Two different monochromators and photon energies were used, with resolution 18 and 7 eV FWHM.  Those data have never been quantitatively analyzed to extract the shakeup and shakeoff spectrum free of instrumental resolution broadening.  That is the objective of the present work.

It would in principle be possible to deconvolve the experimental resolution function from the measured spectra, but the results would be too noisy to be useful.  Instead, we make use of available theoretical and experimental information to construct the salient features of the spectrum, with the theoretically uncertain parameters  (energies, intensities, and shakeoff line shapes) as the fit parameters. Specifically, theory is used to predict: 
\begin{itemize}
    \item The quantum numbers and level ordering of all 2-hole shakeup and shakeoff states,
    \item The relative intensities of shakeup states from a given filled subshell from valence to the continuum edge,
    \item The excitation energies of shakeup states from a given filled subshell from valence to continuum in a hydrogenic approximation,
    \item The shape of the shakeoff continuum excitations with a Levinger distribution \cite{PhysRev.90.11} scaled by a single parameter, 
    \item The total widths of shakeup and shakeoff states as the sum of the widths of the core state and the additional vacancy, and
    \item The statistical weights of spin-orbit partners from their total angular momenta.
\end{itemize}
The objective being the gas-phase spectrum, no plasmon excitations are included.  Only 2-hole final states are considered, as 3-hole states and correlation satellites tend to become less important at the high core-state ionization energies of interest here.  Dense additional structure of that origin can be seen  with low-energy photoionization  \cite{KIKAS1996241}, but it is expected \cite{PhysRevLett.67.2291} to be about an order of magnitude weaker than the 2-hole states in the conversion-line spectrum.

With this information a raw spectral distribution is constructed.  Certain experimental inputs are utilized without adjustment:
\begin{itemize}
    \item The measured widths of vacancy states in Kr and Rb, and
    \item The measured spin-orbit splittings of states in Rb.
\end{itemize}
The spectral distribution is convolved with Gaussian resolutions appropriate to the experimental instrumental widths, and then fit to the experimental data of \cite{PhysRevLett.67.2291} by variation of:
\begin{itemize}
    \item The Gaussian component of instrumental widths,
    \item The amplitudes and energies of shakeup groups, maintaining the hydrogenic spacings and the theoretical shakeup amplitude ratios within each group, and
    \item The amplitude and scale parameter of shakeoff distributions.
\end{itemize}
The fitted spectrum, the uncertainties, and the correlation matrix are the results presented in more detail below.

In Fig.~1 of \cite{PhysRevLett.67.2291},  the data were convolved with a Gaussian to broaden the 18-eV-wide line of the SSRL instruments in order to match the resolution of the Los Alamos spectrometer, about 24 eV.  The variance of the LANL instrumental response function ranged from 85 to 106 eV$^2$ \cite{robertson:1991aa} in 3 data campaigns. Figure~2 of \cite{PhysRevLett.67.2291} shows photoelectron data with undiluted 7-eV FWHM resolution, but the statistical accuracy is lower.  Theoretical calculations give the energies of the satellites quite well with some exceptions but the intensities not so well.  Table \ref{tab:shakestates} 
\begin{table}[htb]
\begin{centering}
\caption{Calculated \cite{PhysRevLett.67.2291} Kr 1s ionization cross
sections and double excitation cross sections
(in percent of the former). For double
ionization (shakeoff) cross sections, only
integrated values are listed.
}
\begin{tabular}{lccc}
\hline
Final state &\multicolumn{1}{c}{Intensity, \%}&\multicolumn{2}{c}{Binding, eV} \\
  & 17025 eV   & \cite{PhysRevA.8.2887, PhysRevLett.67.2291} & \cite{PhysRevLett.57.1566,doi:10.1002/sia.740030412,doi:10.1063/1.435723,tableisotopes}  \\
\hline
$1s^{-1}  \epsilon p $   & 8190 b     &  &  \\
$1s^{-1} 4p^{-1}\ 5p \ \epsilon p$
  &  13.1  & 19.8 & 19.3  \\
$1s^{-1} 4p^{-1}\ 6p \ \epsilon p$
  &   2.5  & {\it 21.7} & 22.9  \\
$1s^{-1} 4p^{-1}\ 7p \ \epsilon p$
  &   0.9  & {\it 22.9} & 24.0  \\
$1s^{-1} 4p^{-1}\ np \ \epsilon p $
  &   1.8  & {\it 23.6} &  \\
$1s^{-1} 4p^{-1} \ \epsilon p \ \epsilon' p $
  &   7.0   & 26.1 & 26.7  \\
Total $4p$ shake&  25.3 &   \\
$1s^{-1} 4s^{-1} \ 5s \ \epsilon p $
  &   2.0  & 36.3 & 35.4  \\
$1s^{-1} 4s^{-1} \ 6s \ \epsilon p $
  &   0.4 & {\it 38.7} & 40.6  \\
$1s^{-1} 4s^{-1} \ 7s \ \epsilon p $
  &   0.1  & {\it 40.2} &  \\
$1s^{-1} 4s^{-1} \ ns \ \epsilon p$  
  &   0.1 & {\it 41.2} &  \\
$1s^{-1} 4s^{-1} \ \epsilon s \ \epsilon' p $
  &   1.3 & 44.3 & 45.2  \\
Total $4s$ shake
&  3.9 &   \\
$1s^{-1} 3d^{-1} \ 4d- nd  \ \epsilon p $
  &  1.3 & 117.0 &  \\
$1s^{-1} 3d^{-1} \epsilon d  \ \epsilon' p $
  &   3.8 & 135.0 & {\it 111.5}  \\
Total $3d$ shake &  5.1 &   \\
$1s^{-1} 3p^{-1} \ 5p- np  \ \epsilon p $
  &  0.1  & 255.4 &   \\
$1s^{-1} 3p^{-1} \epsilon p  \ \epsilon' p $
  &   1.2 & 267.5 & {\it 245.4}  \\
Total $3p$ shake&  1.1 &  1.3 &   \\
$1s^{-1} 3s^{-1} \epsilon s  \ \epsilon' p $
  &  {\it 0.2}  &  & {\it 321}   \\
$1s^{-1} 2p^{-1} \epsilon p  \ \epsilon' p $
   &  {\it 0.3}  &  & {\it 1835}   \\
Total shakeoff   &   13.8   \\
Total shakeup   &   22.3   \\
Total shake     &   36.1   \\
\hline
\end{tabular}
\label{tab:shakestates}
\end{centering}
\end{table}
reproduces Table 1 in \cite{PhysRevLett.67.2291}, adding the binding energies from the RDF theory used in that paper. A related theoretical paper on photoabsorption in Kr \cite{PhysRevA.47.1953} has the same information as \cite{PhysRevLett.67.2291} about the energies of double-hole states (noting, however, that typographical errors exist in their Table II). For the deeply bound $3s^{-1}$ and $2p^{-1}$ states not treated in \cite{PhysRevLett.67.2291}, we use the intensities from the non-relativistic calculation by Carlson and Nestor \cite{PhysRevA.8.2887}, indicated in italics in the Table.  The energies are based on the lines shown in Fig. 1 and 2 of \cite{PhysRevLett.67.2291}, filling in the ones not listed by means of a hydrogenic sequence of energies. Those entries are indicated in italics, with the binding energy for $n'\ge n+1$, \begin{eqnarray}
B_{nln'l}&=&B_{nl\epsilon l}-(B_{nl\epsilon l}-B_{nl(n+1)l})\left(\frac{n+1}{n'}\right)^2.
\end{eqnarray}
Also shown in Table \ref{tab:shakestates} are the binding energies calculated in  RDF and nonrelativistic Hartree-Fock frameworks by Deutsch and Hart \cite{PhysRevLett.57.1566}.  They show that the Rb (Z+1) approximation also works very well ($\lesssim 1$ eV), and we extend their tabulation with entries shown in italics using data from \cite{doi:10.1002/sia.740030412,doi:10.1063/1.435723,tableisotopes}.  

The core state and shakeup states with the quantum numbers $i\equiv1s^{-1} nl^{-1}\ n'l \ \epsilon p$ are given a Lorentzian profile,
\begin{eqnarray}
S(i;E)&=&\frac{A_i\Gamma_i}{2\pi[\Gamma_i^2/4+(E_\gamma-E-B_K-B_i)^2]}, 
\end{eqnarray}
where $\Gamma_i$ is the FWHM of the distribution, $A_i$ the normalization, $E$ is the energy of the ejected core electron, $E_\gamma$ is the initial photon energy, and $B_K$ is the K-shell binding energy.  Recoil effects are not explicitly included.  The width is the sum of the single-particle widths of the core and outer vacancy.  The widths from Table~\ref{tab:lines} are combined, and the results shown in Table~\ref{tab:parameters}. We further augment the spectrum by splitting spin-orbit partners.  The splittings used are those for Rb: 0.69 eV (4p) \cite{PhysRevA.84.053426}, 1.49 eV (3d) \cite{doi:10.1002/sia.740030412}, 8.9 eV (3p) \cite{doi:10.1063/1.435723}, and 60 eV (2p) \cite{tableisotopes}.  The relative intensities of each member of a spin-orbit pair are fixed at the statistical weight $2j+1$.

\begin{figure}[htb]
    \centering
    \includegraphics[width=3.38in]{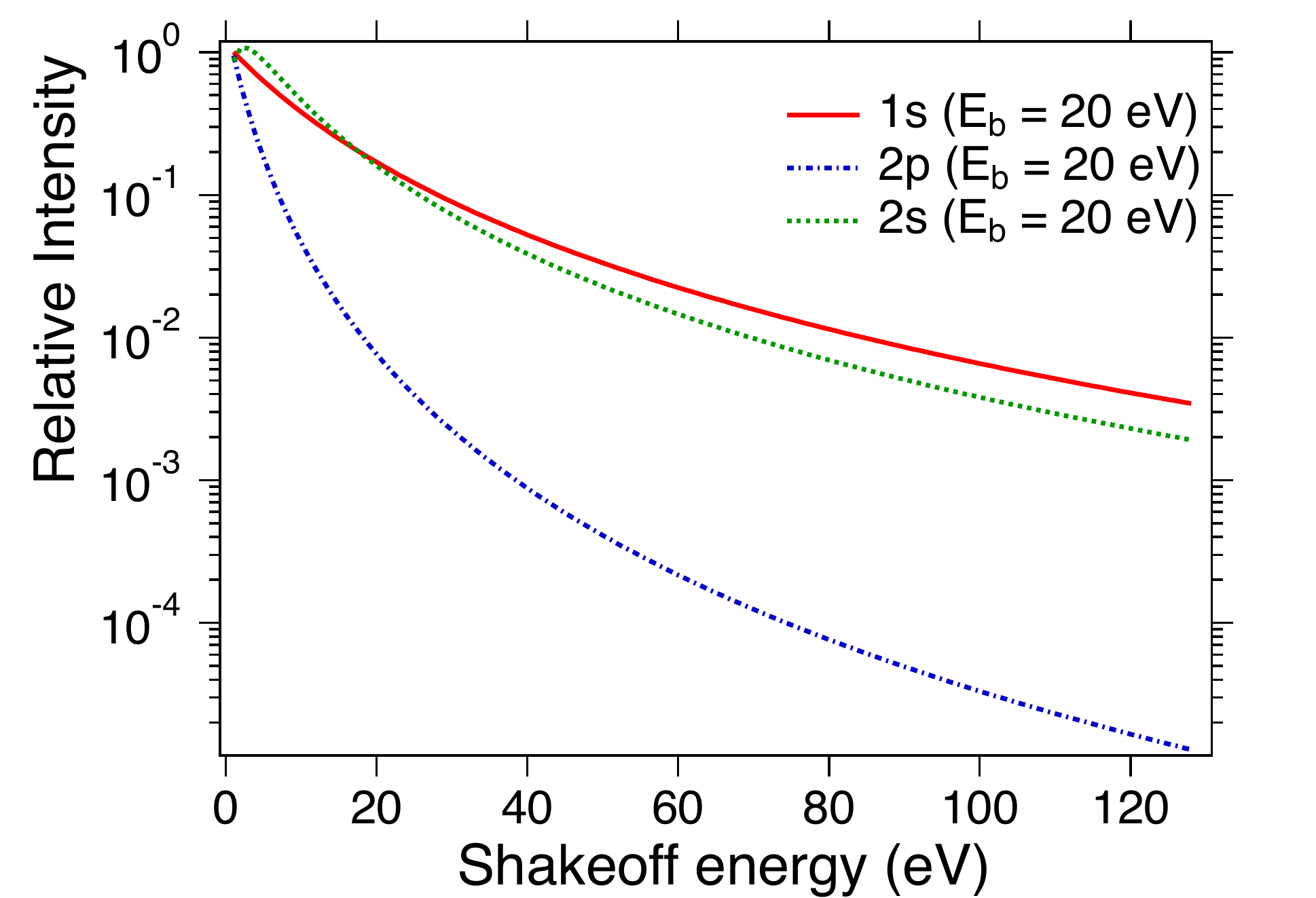}
    \includegraphics[width=3.38in]{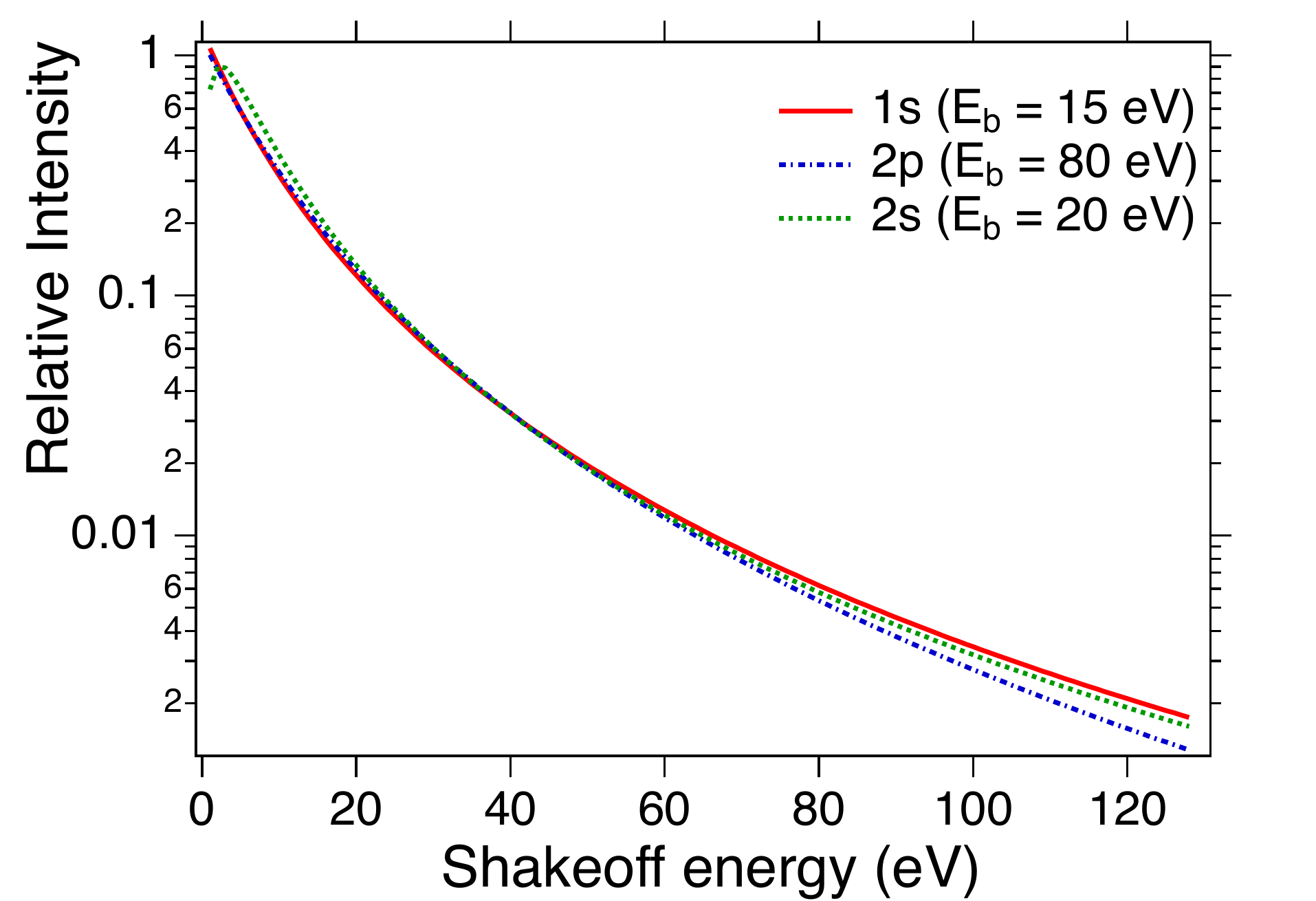}
    \caption{Levinger distributions for 1s, 2s, and 2p shakeoff \cite{PhysRev.90.11}.  Top: $E_b=20$ eV for all. Bottom, $E_b$ chosen to illustrate the approximate scaling of spectral shapes.}
    \label{fig:LevingerRaw}
\end{figure}

Analytic expressions for the shakeoff spectral distributions in hydrogenic atoms were obtained by Levinger \cite{PhysRev.90.11} for emission from three states, 1s, 2s, and 2p:
\begin{widetext}
\begin{eqnarray}
P(1s,\kappa)dW&=&C_{1s}(1-e^{-2\pi \kappa})^{-1}\kappa^8(\kappa^2+1)^{-4}\exp{[-4\kappa\arctan{(1/\kappa)}]}dW \\
P(2s,\kappa)dW&=&C_{2s}(1-e^{-2\pi \kappa})^{-1}\kappa^8(3\kappa^2+4)(\kappa^2+4)^{-6}\exp{[-4\kappa\arctan{(2/\kappa)}]}dW \\
P(2p,\kappa)dW&=&C_{2p}(1-e^{-2\pi \kappa})^{-1}\kappa^{10}(\kappa^2+1)(\kappa^2+4)^{-6}\exp{[-4\kappa\arctan{(2/\kappa)}]}dW
\end{eqnarray}
\end{widetext}
where $\kappa=\sqrt{E_b/W}$, with $W$ the positive kinetic energy of the outgoing shakeoff electron and $E_b$ its  binding energy.  These functions are graphed in Fig.~\ref{fig:LevingerRaw}.
In the upper panel one sees a modest dependence of the spectral shape on principal quantum number $n$, and a more dramatic dependence on $l$.  In the lower panel, an adjustment of $E_b$ is sufficient to produce a common spectral shape.  We therefore use the 1s function for all relevant $nl$ states, and allow both the binding energy and the normalization to be effective (fit) parameters. A similar strategy has been used by Saenz and Froelich \cite{PhysRevC.56.2162} for molecular tritium, but here we are also assuming the scaling behavior persists to larger values of $n$ and $l$. Introducing a leading constant
\begin{eqnarray}
C_{1s}&=&98.2(E_0/E_b)^{1/2}
\end{eqnarray}
reduces correlations between the amplitude and scale factor of the shakeoff distributions.  The intensity normalization is unity when the arbitrary constant $E_0=E_b$. The Levinger distributions use a large-Z approximation, but serve our purpose here.

The shakeoff spectral distribution must be convolved with the Lorentzian width of the 2-hole state in question to produce the final shakeoff distribution,
\begin{multline}
S(i;E) \\ =\int_0^\infty\frac{A_i\Gamma_iP(1s,\kappa)dW}{2\pi[\Gamma_i^2/4+(E_\gamma-E-B_K-B_i-W)^2]}.
\end{multline}
Numerical convolution has been carried out successfully in test fits, but is computationally intensive inside a fitting procedure.  A good approximation that is simpler is to take advantage of the narrow Lorentzian widths in comparison with the typical shakeoff widths.  The convolution of a Lorentzian with a step is an analytic function, suggesting the following form:
\begin{multline}
S(i;E)\\ \simeq A_i\left[\frac{1}{\pi}\arctan\left(\frac{2W}{\Gamma_i}\right)+\frac{1}{2}\right]P(1s,\kappa), 
\end{multline}
wherein $\kappa$ is replaced with $\kappa=\sqrt{E_b/|W|}$ so that $W=E_\gamma-E-B_K-B_i$ can run over all non-zero values.
Since we treat $E_b$ as a fit parameter, in general $E_b\ne B_{nl\epsilon l}$.

With these preliminaries, then, we have a complete listing of the positions of shakeup and shakeoff satellites, and moderate-resolution experimental spectra from which the optimized positions and intensities can be obtained.  The original data are no longer available, and it was necessary to use software to read the points off the plots in \cite{PhysRevLett.67.2291}.  Another source of conversion-line data is the work of Decman and Stoeffl \cite{PhysRevLett.64.2767}, but unfortunately it was not possible to recover analyzable data from the published plots.

\begin{figure}[htb]
    \centering
    \includegraphics[width=3.3in]{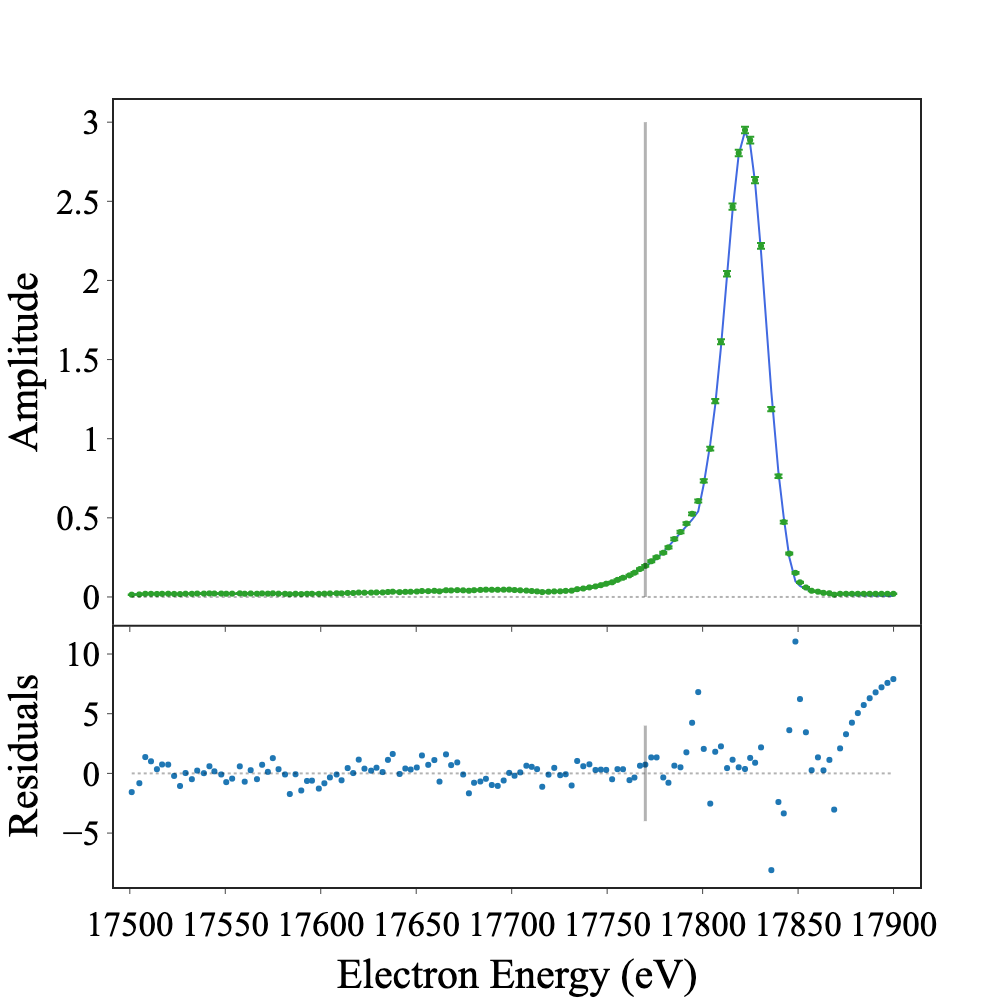}
    \includegraphics[width=3.3in]{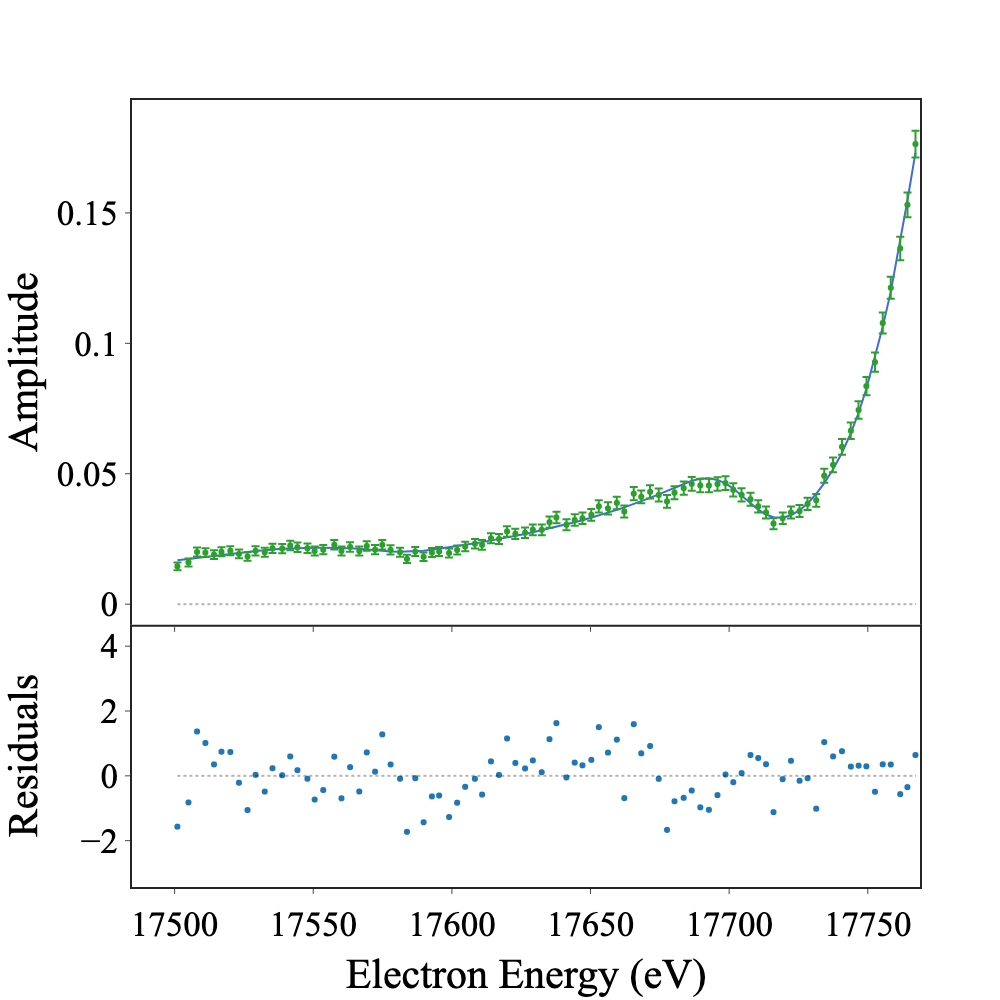}
    \caption{Fits and residuals to the data of Wark {\em et al.} \cite{PhysRevLett.67.2291} from their Fig.~1.  The residuals are in standard deviations, data - fit. The vertical line shows the restricted range for the final fits.}
    \label{fig:fits}
\end{figure}

\begin{figure}[htb]
    \centering
    \includegraphics[width=3.3in]{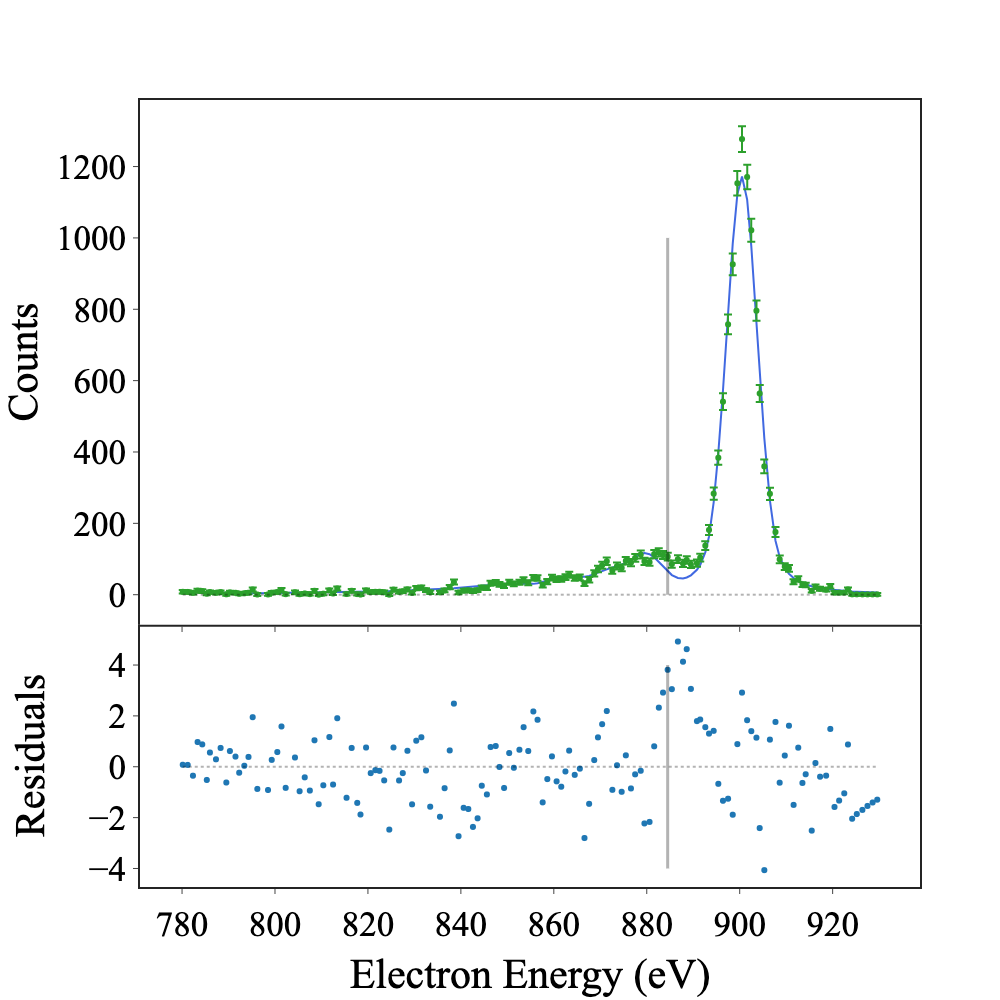}
    \includegraphics[width=3.3in]{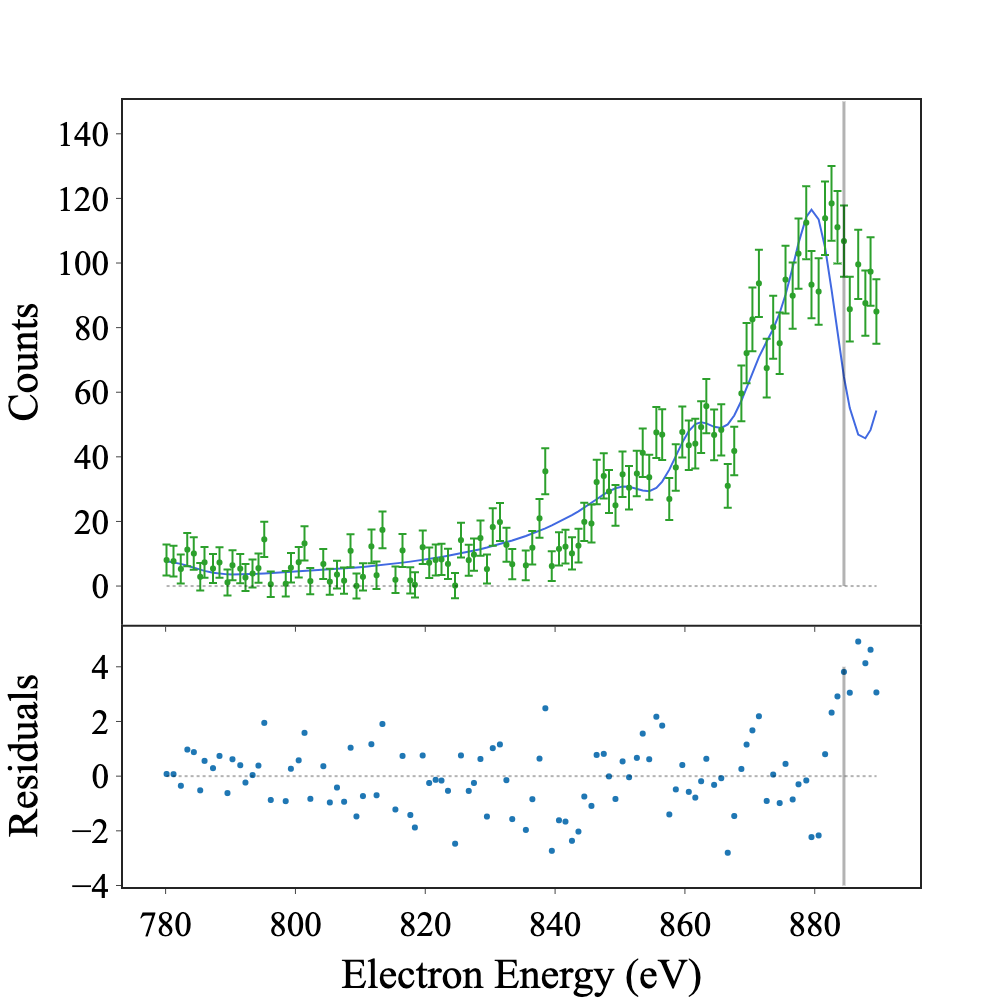}
    \caption{Fits and residuals to the data of Wark {\em et al.} \cite{PhysRevLett.67.2291} from their Fig.~2.  The residuals are in standard deviations, data - fit. The vertical lines show the restricted range for the final fits.}
    \label{fig:fits2}
\end{figure}

To fit the spectrum, we used {\fontfamily{qcr}\selectfont iminuit}  \cite{Ongmongkolkul:2012zz} for its ability to calculate the covariance matrix as well as minimize chi-squared values given constrained tunable parameters in a non-linear setting. The fit strategy is encapsulated in Table~\ref{tab:xtable}.  
\begin{table}[htb]
    \centering
    \caption{Relationship between the fitted parameters with the named ID (e.g. x1) and the states $i$ (see Table~\ref{tab:parameters}) being fitted as a group.  All amplitude and scale factors are constrained to be $\ge 0$. The uncertainties in the top two blocks are as percentages of the core intensity for the summed intensity in that group. Fit parameters x5, x10, and x15 are not used.}
    \begin{tabular}{ccccc}
    \hline \hline
    Parameter ID & \phantom{aa} State $i$\phantom{aa} & \multicolumn{2}{c}{Uncertainty} & Unit \\
    \hline
    \multicolumn{5}{l}{Intensity of shakeoff $\sum A_i$} \\
    x1 & 9, 10 & -0.58  & 0.58  &   \% \\
    x2 & 15 & -1.36 & 1.17 &   \%\\
    x3 & 18, 19 & -0.18 &0.18  &   \%\\
    x4 & 22, 23 & -0.19  &0.43  &   \%\\
    \hline
    \multicolumn{5}{l}{Intensity of shakeup $\sum A_i$} \\
    x6 & 1 -- 8 & -0.47 & 0.47 &   \%\\
    x7 & 11 -- 14 & -1.11 & 0.74  &   \% \\
    x8 & 16, 17 & -0.03   &0.07  &   \% \\
    x9 & 20, 21 &   -0.05 & 0.05 & \%\\
    \hline
    \multicolumn{5}{l}{Binding Energy $B_i$} \\
    x11 & 1 -- 10 & \multicolumn{2}{c}{fixed} & \\
    x12 & 11 -- 15 & -1.3 & 1.3 &   eV\\
    x13 & 16 -- 19 &-0.9 & 0.9 &   eV\\
    x14 & 20 -- 23 & -5 & 5  &   eV\\
    \hline
    \multicolumn{5}{l}{Scale parameter $E_{bi}$} \\
    x16 & 9, 10 & -0.5 & 0.6 &   eV\\
    x17 & 15 &-3.4 & 4.4 &   eV\\
    x18 & 18, 19 & -21 & 24 &  eV\\
    x19 & 22, 23 &-110  &400  &  eV\\
    \hline \hline
    \end{tabular}
    \label{tab:xtable}
\end{table}
The two spectra of \cite{PhysRevLett.67.2291} were fit together to derive a consistent set of shakeup and shakeoff parameters. Points near the high-energy ends (11 in Fig. 1 and 5 in Fig. 2 of \cite{PhysRevLett.67.2291}) were excluded because they could not be accurately read from the plots. The core peaks were fit with a Voigt profile to extract the instrumental width from the known natural width.     The core peaks were found to be at 17822.67(4) and 900.48(4) eV, and the Gaussian instrumental widths  were 23.17(7) and 6.48(12) eV FWHM, respectively.  (The uncertainties are statistical only, and do not include calibration uncertainties.) The fits and residuals are shown in Figs.~\ref{fig:fits} and~\ref{fig:fits2}.  The total $\chi^2$ is 710 for 241 degrees of freedom.  As can be seen from the residual plots, the peak regions contribute non-statistically to $\chi^2$. The final parameters therefore are obtained from fits restricted to  $\le 17770$ eV in Fig.~1 and  $\le884.5$ eV in Fig.~2 because of uncertainty in the instrumental line shape, as discussed below.  Excluding the core peak regions, as indicated in the figures, has little effect on the shakeup and shakeoff parameter values themselves.  The 4p shakeup intensity decreases by one standard deviation and other parameters change by a fraction of a standard deviation. The Gaussian parameters from the full fit are used in the restricted fit.

Fits to the shakeup and shakeoff regions yielded several local $\chi^2$ minima.  They arise from ambiguity in assigning a spectral feature in the data to a corresponding theoretical one.  The lowest $\chi^2$, 200, assigned all the 4p strength to shakeup and none to shakeoff.  That solution was rejected in light of the data of Picard {\em et al.} \cite{picard:1992aa}, which show in the L-line spectra that the shake satellite is centered at 26 eV binding, not 20, and is therefore mainly shakeoff.  Other minima in $\chi^2$ with values 213, 219, 223, and 225 placed the 3d shakeoff edge at 144 eV below the core state, which is inconsistent with Rb photoionization data \cite{doi:10.1002/sia.740030412,doi:10.1063/1.435723}.  The solution consistent with all known independent constraints had a minimum $\chi^2$ of 212. Given the 90+101 data points and 15 tunable parameters, $\chi^2$  per degree of freedom is 1.20. 

As noted above, in the higher-resolution spectrum (Fig.~2 of \cite{PhysRevLett.67.2291}), there are events in a region, the tail of the core peak, that should be devoid of states. The theoretical expectation that there are no 2-hole shakeup states within 20 eV of the core line is supported by the high-resolution data on photoionization of the 3d and 3p states in Kr reported by Eriksson {\it et al.} \cite{Eriksson:1987zz}, although earlier photoionization measurements by Spears {\it et al.} \cite{PhysRevA.9.1603} do show some weak intensity between 10 and 20 eV.  The RDF theory is further supported in predicting the 4p shakeoff edge to be at 26.1 eV, in good accord with the ionization potential for the isoelectronic ion Rb II, 27.28 eV.  The KATRIN internal-conversion data \cite{Altenmuller:2019ddl} cover the region within 15 eV of the 17.8-keV line and find it to be empty, and the Picard {\it et al.} \cite{picard:1992aa} spectra, particularly for the intense, narrow L3 line, show the 20-eV interval to be empty.  Events in that region therefore imply that the instrumental response is not symmetrical and has a more intense low-energy tail than the high-energy one, which is well described by the Voigt profile. For this reason,  the energy of the first 4p shakeup excitation has been fixed at 19.8 eV from the RDF prediction of \cite{PhysRevLett.67.2291} (Table~\ref{tab:shakestates}) because it cannot be reliably determined from this spectrum alone.

Similarly, in the lower-resolution data (Fig.~1 of \cite{PhysRevLett.67.2291}) there is evidence that the peak shape deviates slightly from the Voigt profile. This is not  unexpected, as the theoretical Darwin profile for crystal-diffraction monochromators is not the same as a Voigt profile, and may itself be modified by incidental effects such as heating. 

It may be remarked that a different picture is seen in the Kr {\em threshold} 1s photoexcitation data of Deutsch and Hart \cite{PhysRevLett.57.1566}.  A rich and complex spectrum of weak satellites occupies the excitation region between 12.3 and 19.3 eV binding.  These states consist of correlation and multiparticle satellites, states that would not be excited in the sudden approximation where the ejected electron has much higher kinetic energy than the binding energies. Such states fade to insignificant intensity at the energies imparted by internal conversion in \krm.

\begin{figure*}[htb]
    \centering
    \includegraphics[width=5in]{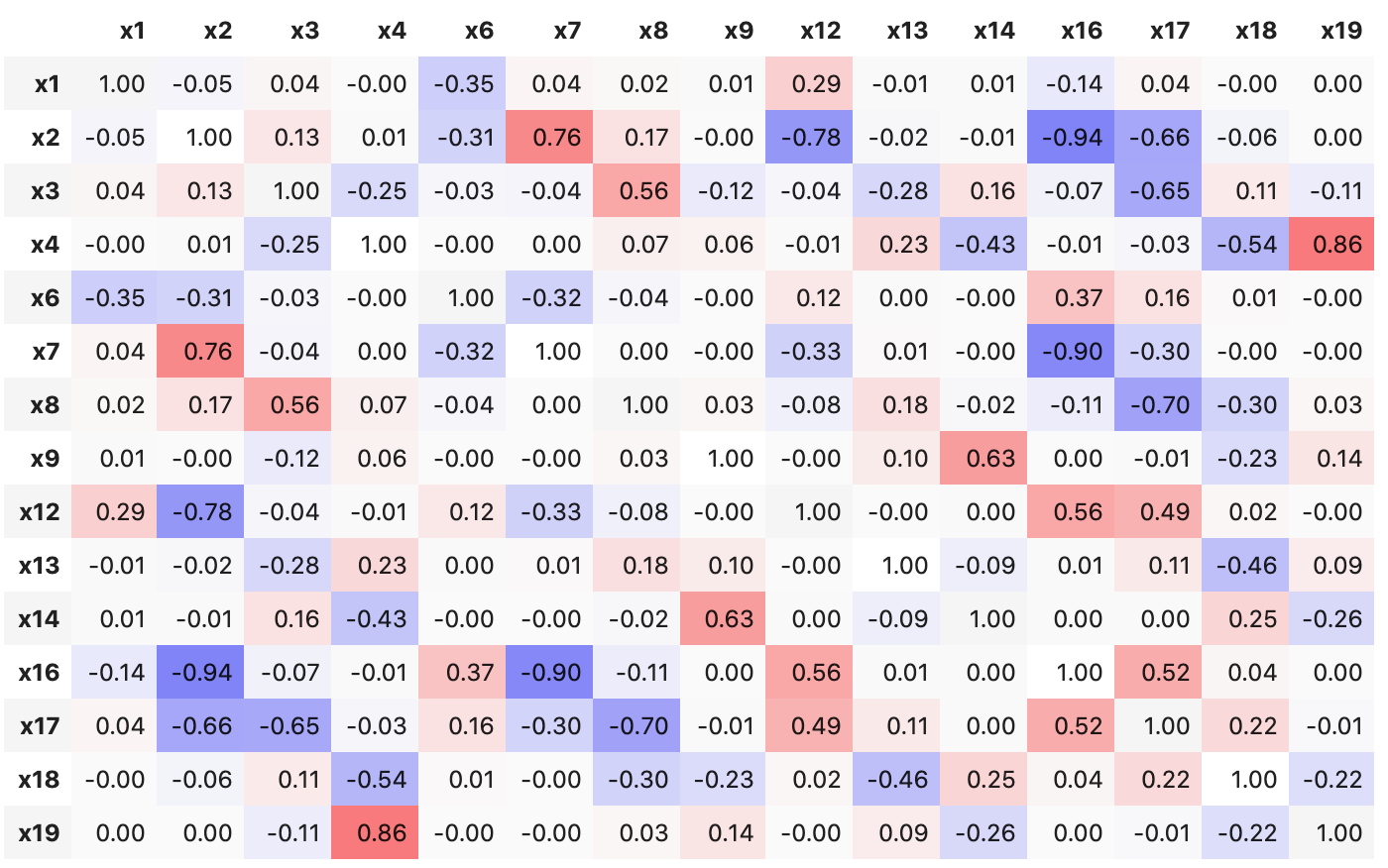}
    \caption{Correlation matrix for the fitted spectrum parameters. Parameters are defined in Table~\ref{tab:xtable}.}
    \label{fig:correlation}
\end{figure*}

\section{Results and conclusions}

\begin{figure*}[htb]
    \centering
    \includegraphics[width=6in]{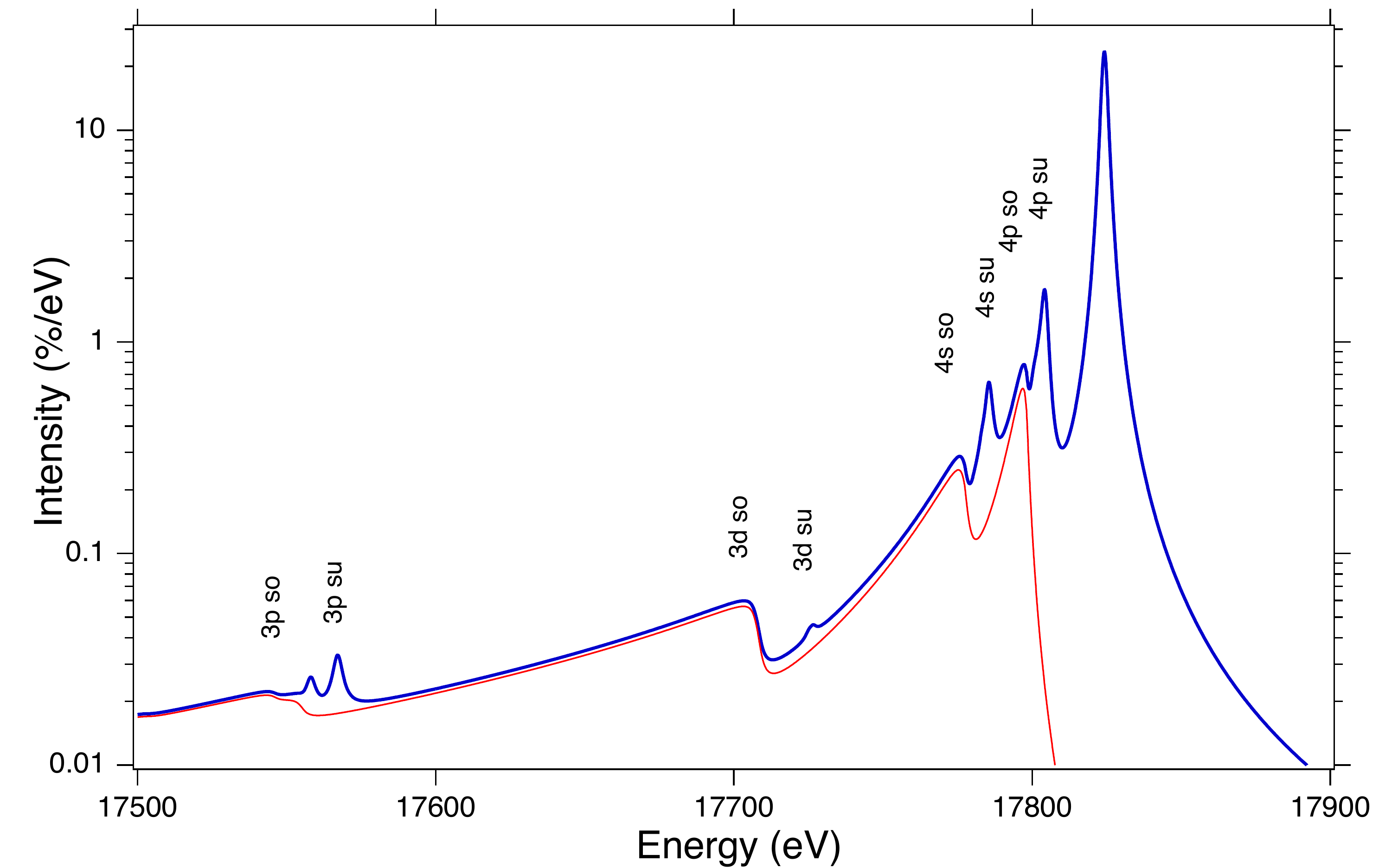}
    \caption{Spectrum of the K-conversion line of \krm\  showing the shakeup (`su') and shakeoff (`so') satellite structure extracted from the calculated line profiles fitted to the data  of \cite{PhysRevLett.67.2291} (upper curve, blue online). The lower curve (red online) shows the shakeoff contribution alone.}
    \label{fig:shakespectrum}
\end{figure*}

In Table~\ref{tab:parameters} are summarized the parameters of the individual components that comprise the full shakeup and shakeoff spectrum of the K-32 line of \krm, and 
\begin{table*}[htb]
\caption{Parameters of Lorentzian and Levinger distributions for the shakeup and shakeoff spectrum of the K-32 line of \krm.  
Uncertainties in the widths are also shown.  For uncertainties in other parameter groups, see Table~\ref{tab:xtable}.
}
\begin{tabular}{llddcc}
\hline
$i$&Final state&\multicolumn{1}{c}{Intensity}&\multicolumn{1}{c}{Binding} &\multicolumn{1}{c}{Width}&Scale\\
&&\multicolumn{1}{c}{$A_i$, \%}& \multicolumn{1}{c}{$B_i$, eV}&\multicolumn{1}{c}{$\Gamma_i$, eV} &$E_{bi}$, eV\\
\hline
0&$1s^{-1} \epsilon p$&100&0&2.70(6)&\\
1&$1s^{-1} 4p_{3/2}^{-1}\ 5p \ \epsilon p$&4.40&19.8&2.70(6)&\\
2&$1s^{-1} 4p_{1/2}^{-1}\ 5p \ \epsilon p$&2.20&20.4&2.70(6)&\\
3&$1s^{-1} 4p_{3/2}^{-1}\ 6p \ \epsilon p$&0.84&21.7&2.70(6)&\\
4&$1s^{-1} 4p_{1/2}^{-1}\ 6p \ \epsilon p$&0.42&22.3&2.70(6)&\\
5&$1s^{-1} 4p_{3/2}^{-1}\ 7p \ \epsilon p$&0.30&22.9&2.70(6)&\\
6&$1s^{-1} 4p_{1/2}^{-1}\ 7p \ \epsilon p$&0.15&23.5&2.70(6)&\\
7&$1s^{-1} 4p_{3/2}^{-1}\ np \ \epsilon p $&0.60&23.6&2.70(6)&\\
8&$1s^{-1} 4p_{1/2}^{-1}\ np \ \epsilon p $&0.30&24.2&2.70(6)&\\
9&$1s^{-1} 4p_{3/2}^{-1} \ \epsilon p \ \epsilon' p $&4.64&26.1&2.70(6)&11.7\\
10&$1s^{-1} 4p_{1/2}^{-1} \ \epsilon p \ \epsilon' p $&2.32&26.7&2.70(6)&11.7\\
11&$1s^{-1} 4s^{-1} \ 5s \ \epsilon p $&2.06&38.7&3.10(7)&\\
12&$1s^{-1} 4s^{-1} \ 6s \ \epsilon p $&0.41&41.1&3.10(7)&\\
13&$1s^{-1} 4s^{-1} \ 7s \ \epsilon p $&0.10&42.6&3.10(7)&\\
14&$1s^{-1} 4s^{-1} \ ns \ \epsilon p$ &0.10&43.6&3.10(7)&\\
15&$1s^{-1} 4s^{-1} \ \epsilon s \ \epsilon' p $&6.63&46.7&3.10(7)&42.6\\
16&$1s^{-1} 3d_{5/2}^{-1} \ 4d- nd  \ \epsilon p $&0.01&97.7&2.77(6)&\\
17&$1s^{-1} 3d_{3/2}^{-1} \ 4d- nd  \ \epsilon p $&0.01&99.1&2.77(6)&\\
18&$1s^{-1} 3d_{5/2}^{-1} \epsilon d  \ \epsilon' p $&4.43&115.7&2.77(6)&304\\
19&$1s^{-1} 3d_{3/2}^{-1} \epsilon d  \ \epsilon' p $&2.95&117.1&2.77(6)&304\\
20&$1s^{-1} 3p_{3/2}^{-1} \ 5p- np  \ \epsilon p $&0.09&257.1&4.02(6)&\\
21&$1s^{-1} 3p_{1/2}^{-1} \ 5p- np  \ \epsilon p $&0.04&266.1&3.93(9)&\\
22&$1s^{-1} 3p_{3/2}^{-1} \epsilon p  \ \epsilon' p $&0.77&269.2&4.02(6)&254\\
23&$1s^{-1} 3p_{1/2}^{-1} \epsilon p  \ \epsilon' p $&0.37&278.2&3.93(9)&254\\
24&$1s^{-1} 3s^{-1} \epsilon s  \ \epsilon' p $&0.2&321&6.20(40)&\\
25&$1s^{-1} 2p_{3/2}^{-1} \epsilon p  \ \epsilon' p $&0.2&1835&3.81(6)&\\
26&$1s^{-1} 2p_{1/2}^{-1} \epsilon p  \ \epsilon' p $&0.1&1895&3.87(9)&\\
\hline
\end{tabular}
\label{tab:parameters}
\end{table*}
Fig.~\ref{fig:shakespectrum} shows the spectrum of the K-conversion line using the fitted parameters listed in the table.

The parameter uncertainties are included in Table~\ref{tab:xtable}. The uncertainties were obtained with the {\fontfamily{qcr}\selectfont  minos} routine of {\fontfamily{qcr}\selectfont iminuit} that searches each parameter in turn for the change that increases $\chi^2$ by one,  marginalizing over the others.  In general the uncertainties are  asymmetric.  The correlation matrix determined from the Hessian is presented in Fig.~\ref{fig:correlation}. The largest element is -0.94 between the 4s shakeoff amplitude and the scale factor of 4p shakeoff.

The total intensity in shakeup and shakeoff below the core peak is found to be 34.7\% of the core peak, in  good agreement with the theoretical calculation (see Table \ref{tab:shakestates}).  However, the fits  confirm the observation \cite{PhysRevLett.67.2291} that the RDF theory tends to overemphasize shakeup at the cost of shakeoff, even while conserving the total probability, for reasons that are at present not known.  Moreover, the RDF calculations of \cite{PhysRevLett.67.2291} overbind the 3d states by about 19 eV.

This work was motivated by the development of the cyclotron radiation emission spectroscopy (CRES) method, exemplified by Project 8 \cite{Asner:2014cwa}.  Electrons emitted from a radioactive gas ($^3$H and \krm\ particularly) are trapped in a magnetic trap for a precise measurement of their cyclotron frequencies and, hence, energies. Electrons escape by scattering from background gas atoms, and several scatters are typically needed to eject the electron.  The trap thus contains both scattered and unscattered electrons, an effect that must be taken into account in determining the instrumental response.  The \krm\  lines provide an ideal testbed for this determination, but the shakeup and shakeoff satellites occupy the same region of the spectrum as scattered electrons, and therefore must be quantitatively treated. 

The KATRIN experiment now in operation \cite{PhysRevLett.123.221802} does not use \krm\ for a direct determination of resolution or scattering; an electron gun is used.  The isotope is, however, brought to bear on a number of systematic tests either alone or mixed with tritium (see, for illustration, \cite{Belesev:2008zz}).  Even so, the satellite structure of the lines plays little role in the interpretation of those tests, and it is not expected that the results reported here will influence the KATRIN program greatly.   KATRIN's high statistical and systematic precision takes \krm\ out of the list of contributions to Eq.~\ref{eq:variance}. 

In summary, the shakeup and shakeoff spectrum derived in this work is intended to serve as an improved prediction of the extended shape of the 17.8-keV internal conversion line of \krm.  The spectrum provided will find utility in  currently running and planned tritium beta decay neutrino mass experiments, such as Project 8 \cite{Asner:2014cwa} and, possibly, KATRIN \cite{Angrik:2005ep,Altenmuller:2019ddl}.  The spectrum also provides a baseline for further theoretical work.  New high-resolution measurements of this spectrum by both internal conversion and photoionization are well within technical reach~\cite{PhysRevB.64.045109} and are  encouraged.

We gratefully acknowledge discussions with Christine Claessens and Gerald Seidler.  This material is based upon work supported by the U.S. Department of Energy Office of Science, Office of Nuclear Physics under Award Number DE-FG02-97ER41020.

\bibliography{krypton}

\bibliographystyle{apsrev}
\end{document}